\documentstyle[11pt,epsfig]{article}
\newcommand{\be}{\begin{equation}}
\newcommand{\ee}{\end{equation}}
\newcommand{\ba}{\begin{eqnarray}} 
\newcommand{\ea}{\end{eqnarray}}
\begin{document}

\hfill{Dec. 16, 1999} 
\begin{center}

{\Large \bf  Evolution of a Rotating  Black Hole with 
a Magnetized Accretion Disk}

\vskip 1cm 

{\bf Hyun Kyu Lee and Hui-Kyung Kim

Department of Physics, Hanyang University, 
Seoul 133-791}

\end{center}

\vskip 2cm

\vskip 0.5cm
\indent The effect of an accretion disk on the Blandford-Znajek 
process and the evolution of a black hole are discussed
using a simplified system for the black 
hole - accretion disk in which the accretion rate is supposed to be dominated
 by the strong
magnetic field on the disk.  The evolutions of the mass and the angular momentum of the 
black hole are formulated and discussed with numerical calculations.   


\section{Introduction}

The system of a rotating black hole - magnetized accretion disk 
is an interesting object  where the rotational energy of 
the black hole can be extracted along the field lines supported 
by the accretion disk\cite{bf, TPM}.  
Recently, there has been increasing interest in 
the possibility that such a system with a stellar black hole 
can be the central engine \cite{lwb} for powering gamma - ray bursts\cite{piran}
via the Blandford-Znajek process originally 
proposed for AGN.     The Blandford-Znajek power for a black hole with 
mass $M$ and angular momentum $J (=\tilde{a} M^2)$ is 
given by 
\ba
P_{BZ} =\frac{1}{4} \tilde{a}^{2} M^{2} B_{H}^{2} f(h) , 
\ea
which  is estimated to be
consistent with the observed cosmological gamma - ray bursts
provided
a strong magnetic field $B_H \sim 10^{15}$G is on the black hole.  

Since the power depends on the angular momentum parameter $\tilde{a}$ and 
the black hole mass $M$,  the evolution of the black hole during the 
powering process may also be important for the time evolution of the power,
which, in principle, is related to the complicated  temporal 
structure of the gamma -  ray 
spectrum.  In the simple model\cite{simon} proposed to explain the temporal 
structure,
one of the important parameters is the evolution of the power in time.
 
The evolution 
of a rotating black hole in the Blandford-Znajek process, 
without the effects of an accretion disk
but with a given magnetic field on the horizon,
has been discussed 
in detail by Okamoto \cite{ok}.    While earlier discussions on the 
effects   
of  the accretion disk can be found in Park and Vishniac\cite{park},
the effect of the accreting materials on the rotation of the 
black hole has continued to be an interesting subject\cite{th, gammie}.  

In this paper, we will discuss  the effects of 
an  accretion disk, in  which the magnetic torque is supposed
to dominate the accretion rate\cite{lwb},  on black hole evolution. 
The magnetic field on the disk determines not only the accretion rates 
of energy and angular
momentum into the black hole
but also the
rates of energy and angular momentum losses from the black hole since 
it 
is related to the magnetic field on the
 horizon.
We suppose a very simple -  minded model for a black hole - accretion
disk, in which 
the transition region is simplified to a region in which nothing happens
(no magnetic coupling 
between the horizon and the disk, for example) except
the flow of accreting matter. 
The accretion disk is assumed to satisfy the condition 
for the axisymmetric solution  proposed 
by Blandford\cite{blandford}
in 1976 for an accretion disk in a 
force-free magnetosphere.  Then, we can find  an analytic relation 
between the  magnetic fields on the horizon and 
the inner edge of the accretion disk.
However, since there is insufficient information on the
origin or the strength of  the disk magnetic field, 
we take it as a 
parameter for the numerical calculations.    
  Although the transition region between the
horizon and the inner edge of the accretion disk is quite complicated, we try to  infer some of the  qualitative features of 
 black - hole
evolution from this simplified analysis.
 
\section{Blandford-Znajek power and accretion rate}

We consider the simplified system of a black hole - accretion disk.  The 
environment around the system is assumed to be a force-free magnetosphere
with axial symmetry. 
The accretion disk is known to be 
a very complicated object compared to the  black hole which is mathematically 
well defined. Provided with the mass ($M$), the angular momentum
($J$), and the charge ($Q$) 
of a black hole, 
we can calculate its various physical quantities:
 for example, the horizon radius ($r_H$) 
and the stable orbit of a test particle.  However, the 
accretion disk is a rather  complicated object with a number of physical parameters
which define the accretion disk\cite{pringle}.  
Among 
them are the shape
parameters (thickness of the disk, for example), the density, the
pressure, the temperature, and the viscosity parameter ($\alpha$). 
While the magnetic field on the horizon 
is assumed to be well ordered \cite{TPM}, 
the magnetic field on the disk can be decomposed into 
 ordered and disordered  
parts. For an accretion disk where the ordered magnetic field dominates  
the disordered magnetic fields, we adopt axial symmetric
and the steady - state solution obtained by Blandford for a force-free magnetosphere \cite{blandford}. 

The outward current density along the disk, $J_{\tilde{\omega}}$, is 
proportional
 to the toroidal component of the magnetic field on
the disk, $B_{\phi}$:
\ba
 2 \pi J_{\tilde{\omega}} = -  B_{\phi}\label{jr}
\ea
where 
cylindrical coordinates $(\tilde{\omega},\phi,z)$ are used. 
The $z-$direction is chosen to be 
perpendicular to the disk. $\tilde{\omega}$ is the cylindrical radius from the
origin.  

The torque exerted by an  annular
ring with width $\Delta r$ \footnote{For a nonrelativistic   discussion
      of the disk, $\tilde{\omega}$ is replaced by a radial 
distance $r$ on the disk.}
 due to
the Lorentz force is given by
\ba
\Delta T = - r \, 2\pi r J_r \, B_z \Delta r.
\label{deltatd}
\ea
For a steady  - state accretion disk with a surface density $\Sigma$,
 angular momentum conservation
can be written as
\ba
\Delta T = \Sigma v_r \frac{\partial(r^2 \Omega)}{\partial r} 
2\pi r \Delta r =
\dot{M}_+ \frac{\partial(r^2 \Omega)}{\partial r}  \Delta r,  
\label{deltaac}
\ea
where the torque due to the shear force of differential rotation is assumed
 to be negligible\cite{lwb}. The accretion rate in the 
above equation is defined by
\be
\dot{M}_+ = \Sigma v_r 2\pi r . \label{mdot}
\ee

 From Eqs.(\ref{deltatd}) and (\ref{deltaac}), using the Keplerian 
angular velocity,
 $\partial(r^2\Omega)/ \partial r = r\Omega_{disk}/2$,
we get 
\ba
\dot{M}_+ = 2rB_{\phi}B_z/\Omega_{disk}. \label{mdot}
\ea
Adopting  the axisymmetric solution\cite{blandford},
\ba
B_{\phi} = 2r\Omega_{disk} B_z/c, \label{bbz}
\ea
we get the accretion rate given by
\ba
\dot{M}_+ = 4 \, r^2 B_z^2/c. \label{mdot1}
\ea

In this simple schematic  model, the  poloidal current flowing 
into the horizon is responsible for the toroidal magnetic field on the horizon 
and on the inner edge of  the accretion disk. 
If there  is no current source 
or sink in the transition region, current conservation  implies 
 that the total current flow, $I$,  onto the black hole
should go into the inner edge ($r_{in}$) of the accretion disk. Then,  
Ampere's law implies
\begin{equation}
2M B_{\phi}^H(\theta=\pi/2) = 4\pi I =\tilde{\omega}(r_{in})
B^{disk}_{\phi}(r_{in}).
\end{equation}
On the other hand,  from the boundary conditions on the horizon\cite{TPM} 
in the optimal case\cite{lwb},
\begin{equation}
B_{\phi}^H = -\Omega_H M B_H,
\end{equation}
and on an accretion disk\cite{blandford} with angular
velocity $\Omega_D$,
\begin{equation}
 B_{\phi}^{disk} =
-2 \Omega_D r B_z ,
\end{equation}
we get the relation between the poloidal magnetic fields on the horizon 
and the disk's inner edge :
\begin{equation}
\frac{B_z(r_{in})}{B_H} =  \sqrt{\frac{GM}{r_{in}c^2}}
\frac{\tilde{a}}{2}\frac{G M}{r_Hc^2} \frac{r_{in}}{\tilde{\omega}}
 \label{bhd} ,
\end{equation}
which implies that the poloidal magnetic field on the horizon is larger than 
that on the inner edge of the disk\footnote{The magnetic braking power from 
the disk in connection with the Blandford-Znajek power from the black hole
 has been discussed recently by Li\cite{li} and by Lee, Brown, and 
Wijers\cite{lbw}.}. A similar observation  has been discussed 
by Ghosh and Abramowicz\cite{ghosh} by using the field configuration motivated 
by MacDonald\cite{mac}.  However, it is not clear yet whether these field
configurations can be realized by the conventional accretion 
dynamo\cite{lop}. 

Using Eqs.(\ref{mdot}) and (\ref{bhd}), we can write the accretion rate in terms 
of the magnetic field on the horizon,
\ba
\dot{M}_+ = \frac{M r^3_{in}}{\tilde{\omega}^2r_H^2} M^2 \tilde{a}^2 B_H^2 
. \label{M0}
\ea
The Blandford-Znajek power is given by\cite{lwb}
\ba
P_{BZ} &=& \frac{1}{4} \tilde{a}^2 M^2 B_H^2 f(h) ,
\ea
which can be written in terms of the accretion rate in Eq. (\ref{M0}) as 
\ba
P_{BZ} =  \frac{1}{4} \frac{\tilde{\omega}^2r_H^2}{M r^3_{in}} 
            \dot{M}_+ f(h),\label{pbzm}
\ea
where
\ba
f(h) &=& \frac{1 + h^{2}}{h^{2}}[(h + \frac{1}{h}) \arctan{h} - 1]
, \,\, \\
h &=& \frac{ \tilde{a}}{H}, \,\,\,  H=1 + \sqrt{1 - \tilde{a}^2}. 
\ea
The horizon radius is $ r_H = HM $.       

\section{Energy and angular momentum accretions into a black hole}

The physics in the transition region may  be quite complicated such that 
it is not certain how much energy and angular momentum can be accreted 
into the black hole horizon. It depends on the magnetic field\cite{gammie}
 and also 
on the particular types of accretions\cite{yi}.   In this work with a simple 
schematic model of a black hole and an accretion disk,  it is
assumed that the inner edge\cite{abramowicz} of the accretion disk is the 
last stable orbit\cite{shapiro} defined  by 
\ba
r_{in} =  \,  Z M,
\ea
where
\ba 
Z &=& 3 + Z_{2} - \sqrt{(3 - Z_{1})(3 + Z_{1} + 2 Z_{2})},\\  
Z_{1} &=& 1 +  (1 -   \tilde{a}^{2})^{\frac{1}{3}}
[(1 + \tilde{a})^{\frac{1}{3}}  + (1   -
\tilde{a})^{\frac{1}{3}}],\\
Z_{2} &=& \sqrt{3 \tilde{a}^{2} + Z_{1}^{2}}.
\ea
It is assumed that the energy of a particle in the last stable orbit 
is accreted into the horizon without any radiation outward. The specific 
energy $\tilde{E}$ (energy per unit rest mass, $E=m\tilde{E}$) 
at $r_{in}$ is given by
\ba
\tilde{E} 
 =  \frac{Z^{2}-2Z + \tilde{a} \sqrt{Z}}{Z \sqrt{Z^{2}-3 Z + 2 \tilde{a}
 \sqrt{Z}}}
\ea
One can see that the accreted energy becomes smaller 
if the black hole is rotating: 
$\tilde{E}(\tilde{a}=0) = 0.94 \, \rightarrow \, \tilde{E}(\tilde{a}=1)
 = 0.58$. The rest 
of the specific energy is assumed to be dissipated during the accreting
 processes.  It is interesting to note that for advection - dominated 
accretion\cite{yi}, the sepcific energy accreted  should be taken to be larger 
than $\tilde{E}$.

The specific angular momentum ($l=m\tilde{l}$) at the last stable orbit, 
which is carried into the horizon, is given by
\ba
\tilde{l}
 =  M {\cal Z}_{l}(\tilde{a}), \label{l0}
\ea
where
\be
{\cal Z}_{l}(\tilde{a})=\frac{Z^{2}-2 \tilde{a} \sqrt{Z} 
+ \tilde{a}^{2}}{\sqrt{Z(Z^2 -3Z + 2\tilde{a}\sqrt{Z})}}.
\ee

In this simplified work, we assume that there is nothing 
particular between the horizon and the inner endge of the disk
execpt the accretion flow, although  recent works\cite{gammie}
 show that 
 energy and angular momentum extraction from a black hole 
onto a disk is  possible, which makes $\tilde{E}$ and 
$\tilde{l}$  smaller. 

\section{ Evolution of a black hole}

The evolution rates  of  the black hole mass ($\dot{M})$ and the angular 
momentum ($\dot{J}$)
are determined both  by 
the accreted energy and the angular momentum,
which increase the mass and the angular momentum, and by the  
Blandford-Znajek power in the opposite 
direction. 
From  energy conservation, we get 
\ba
\dot{M} &=&  - P_{BZ} + \dot{M}_{+} \tilde{E} \\
& = & ( \tilde{E}- \tilde{P}_E) \dot{M}_{+}, \label{mbdot}
\ea
where 
\ba
P_{BZ} &=& \dot{M}_+ \tilde{P}_E, \\
\tilde{P}_E &=&  \frac{f(h)H^2}{4Z} [1 +
(\tilde{a}/Z)^2(1+ 2/Z)]. \label{ptilde}
\ea

Angular momentum conservation leads to 
\ba
\dot{J} &=& - \frac{P_{BZ}}{ \Omega_{F}}+ \dot{M}_{+} \tilde{l}.
\ea
The first term on the right hand side is the angular momentum loss rate
due to the Blandford-Znajek process, and the second term is the accreted angular 
momentum from the disk. Equation(29) can be written as
\ba
\dot{J} &=& M \dot{M}_{+}({\cal Z}_{l}- \tilde{P}_l),
\ea
where 
\ba
\tilde{P}_l =  \frac{fH^3}{\tilde{a}Z}  [1 +
(\tilde{a}/Z)^2(1+ 2/Z)]\label{ptildl}.
\ea

The evolution of the angular momentum parameter
$\tilde{a}$ can be obtained from the angular momentum evolution  
as 
\ba
 \frac{ \dot{J}}{J} &=& \frac{ \dot{a}}{a}+ \frac{ \dot{M}}{M}
\ea
or
\ba
 \dot{J} &=& (2 \frac{ \dot{M}}{M} \tilde{a}+ \dot{ \tilde{a}})M^{2}.
\ea
Then, 
using Eq. (\ref{mbdot}), we get  
\ba
\dot{ \tilde{a}} =  \frac{ \dot{M_+}}{M}[({\cal Z}_{l}-\tilde{P}_l) 
-2\tilde{a}(\tilde{E} - \tilde{P}_E)], \label{adot}
\ea
where the accretion rate is given by
\be
\dot{M}_+ = \frac{Z}{H^2 [1+(\frac{\tilde{a}}{Z})^2(1+\frac{2}{Z})]}
M^2 \tilde{a}^2 B_H^2.\label{mdot1}
\ee

Since the system of a black hole - accretion disk we are considering for 
the central engine of the gamma - ray bursts is supposed to emerge in the final 
stage of  
the binary - merging processes\cite{merge} 
like neutron star - neutron star and  
black hole - neutron star, the mass of the accretion disk should   
not be much greater 
than the solar mass and the lifetime ($\tau$) of 
an accretion disk with appreciable 
pressure for a strong magnetic field cannot be of 
an astronomical scale. From the large accretion rate due to the 
strong magnetic
 field,  is assumed to be less than a few thousand seconds\cite{lwb}.
Since the presence of an accretion disk with an appreciable pressure or magnetic
 field is essential for the magnetic field on the horizon, the evolution
of the accretion disk (lifetime with appreciable pressure in it) might be
also responsible for the evolutions of the magnetic field on the disk and  
$B_H$. 

For the numerical calculations
in this work, 
we tried to incorporate the effects of mass loss from the disk into the 
the time dependence of the magnetic field in the following form : 
\be
B_H^2 = B_H^2(0)D(t),
\ee
where we take
\ba
 D(t) = 1 - (\int^t_0 \dot{M}_+)/M_{\odot}.
\ea   
We take $D(t)$ to be vanishing as the total accreting mass becomes
a solar mass, which is supposed be a typical mass for an  accretion
disk emerging out of binary - merging processes.

\section{ Results and discussion}

The evolution of the black hole is calculated with $M(0) = 3M_{\odot}$
 and $B_H(0) = 10^{15}$G and is shown in Fig. 1 for mass ($M$) and 
in Fig. 2
 for angular momentum ($\tilde{a}$).  When we take the initial angular 
momentum parameter  to be 
 $\tilde{a}(0) = 1$,  the mass of the 
black hole increases up to $\sim 3.6M_{\odot}$. Compared to the
initial mass sum of $4M_{\odot}$, $\sim 10\%$ of the rest - mass energy of the
system is extracted by the Blandford-Znajek process, which is about
the same fraction  of energy that can be taken from the 
black hole only\cite{lwb}.  The interesting observation is that the 
rotation of the black hole does not stop, $\tilde{a} \sim 0.82$, even after
the disk disappears (equivalently, no magnetic field on the black hole or no 
Blandford-Znajek process). 
For a system of only a black hole with a constant magnetic field on it, the
angular momentum decreases rapidly within a few thousand seconds, and in the
 optimal process, the fraction of rotational energy extracted is about 
9\%\cite{lwb}. This implies that the energy extracted from the black hole 
in the black hole - accretion disk system   
is not only from the black hole's initial rotational energy but also 
 from the energy accreted  from the disk.
The Blandford-Znajek power shown 
in Fig. 3 drops very 
rapidly within $\sim 1000$s, as expected, and the pattern of the evolution
with time is similar to the case with the black hole alone\cite{ok}.  
   
In this paper, we formulated the evolutions of the black hole mass and angular 
momentum, taking into account the flow of accreting matter from a strongly 
magnetized disk. Adopting the axisymmetric solution suggested by Blandford,
we got analytic fomulae for the evolutions and could see that the angular momentum
 accreted from the disk 
is comparable to the amount of  angular momentum extracted by the
Blandford-Znajek process. a  numerical calculation  demonstrated
that the accreted angular momentum 
was  large enough to keep the black hole rotating rapidly while
the Blandford-Znajek 
process was  working. 

Since we considered a very simplified situation, there are several issues
to be discussed in the future.  For  the numerical calculation, we
assumed a specific time dependence of the 
magnetic field on the disk,   which was designed to  be vanishing as 
the disk mass decreased.  Although we think the general feature of 
this assumption 
may not be totally wrong, at least, for the isolated 
 black hole - accretion 
disk system,  we need more detailed  analysis to find  
the relation between the magnetic field and the disk properties.  
Also, for a realistic accretion disk, we should   consider the effects of 
the disordered magnetic field, the turbulence, and/or  advection - dominated flow\cite{yi}.  It should be also noted that there are several 
works\cite{lop, park2} demonstrating some difficulties in building up 
a strong  poloidal 
magnetic feld on the disk by means of conventional accretion and dynamo
processes.  
In this work, we have not taken into account the  
recent observations that negative energy flow out to the disk from the 
black hole along the accreting flow is possible when there is a strong 
enough magnetic field in the transition region\cite{gammie}.  
That, together with a relativistic generalization of the Blandford solution and the 
accretion formulae used in this work, would be an interesting problem
for future work.   

\section{acknowledgments}
We would like to thank Insu Yi for useful discussions. 
This work is supproted in part by the Ministry of Education
(BSRI-98-2441), a  Hanyang Uiversity grant in 1999,
and the Korea Science and Engineering Foundation (Grant No. 1999-2-112-003-5).


\newpage

\hspace{5cm} Caption  \\ \\

Figure 1. The evolutions of a black hole with  $M(0) =
    3M_{\odot}$  and  $\tilde{a}(0) = 1$ \\ 

\hspace{1.5cm} for a black hole only(solid
    line) and for a  black hole - accretion \\

\hspace{1.5cm}  disk system(dotted line).\\ \\ \\

Figure 2. The evolution of the angular momentum in units of $M^2$.\\ \\ \\

Figure 3. The time dependences of the Blandford-Znajek power.\\

\newpage
\begin{figure}[htbp]
  \begin{center}
    \leavevmode \epsfig{file=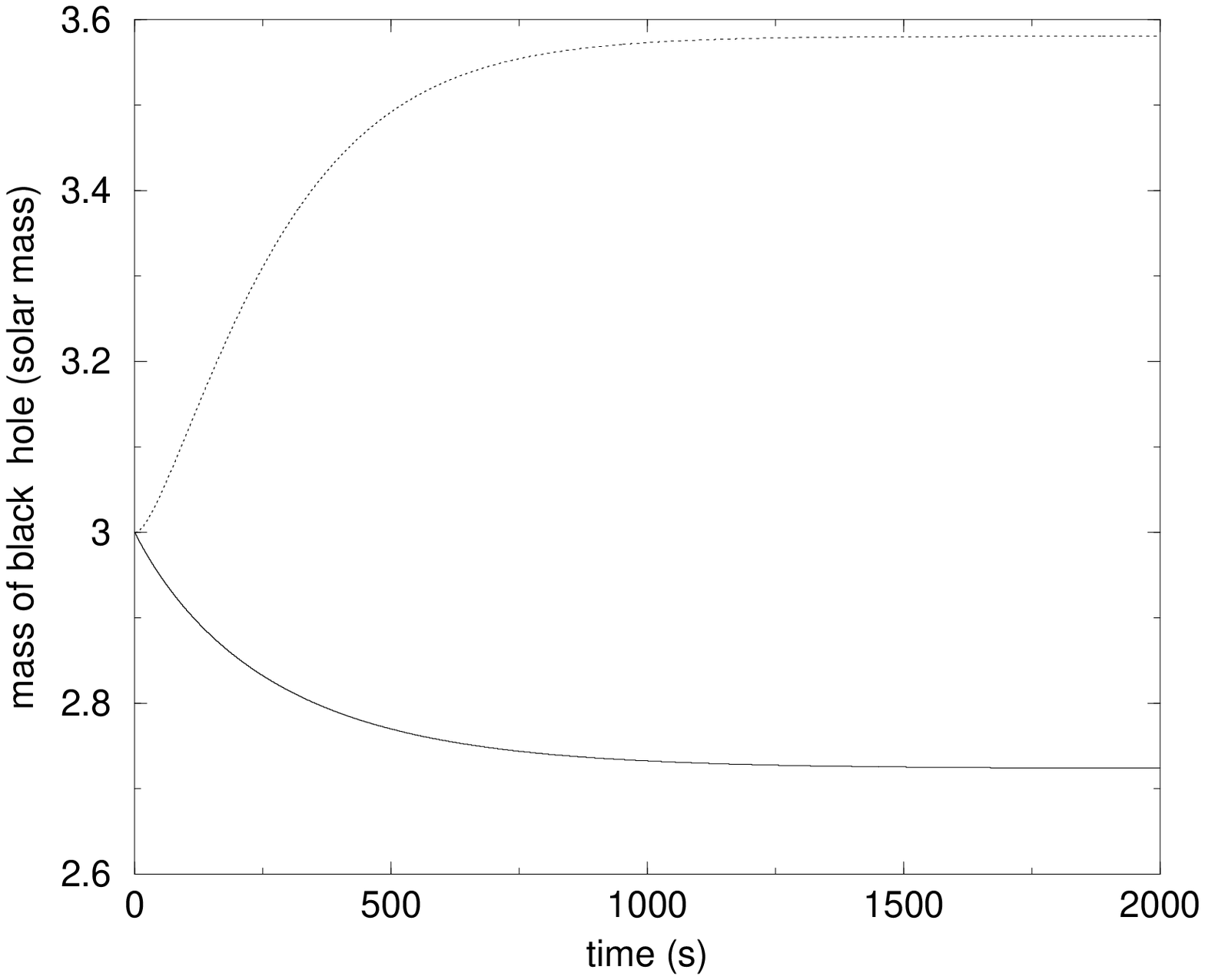, height=11cm}
    \caption{}
    \label{fig1}
  \end{center}
\end{figure}

\newpage
\begin{figure}[htbp]
  \begin{center}
    \leavevmode \epsfig{file=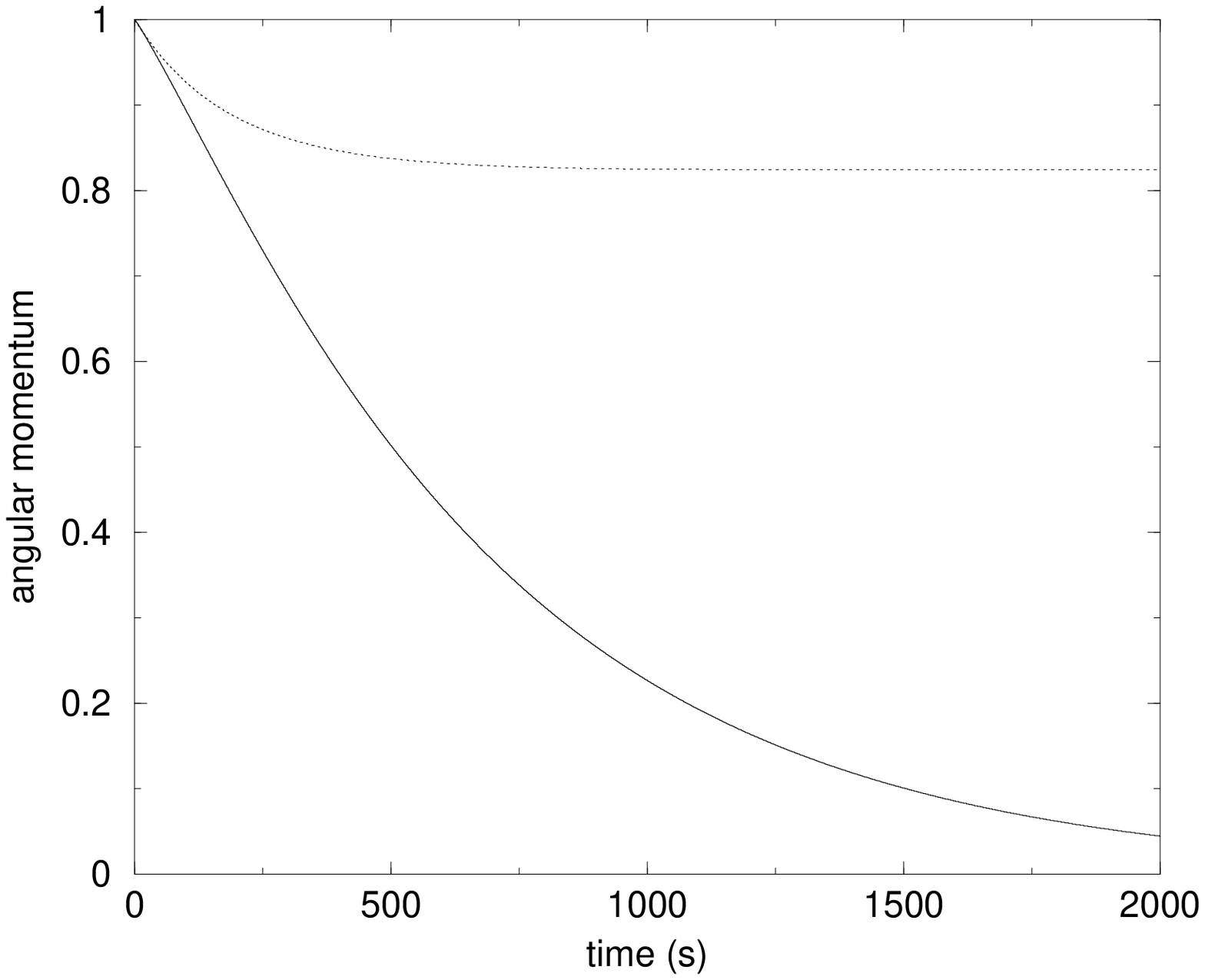, height=11cm}
    \caption{}
    \label{fig2}
  \end{center}
\end{figure}

\newpage
\begin{figure}[htbp]
  \begin{center}
    \leavevmode \epsfig{file=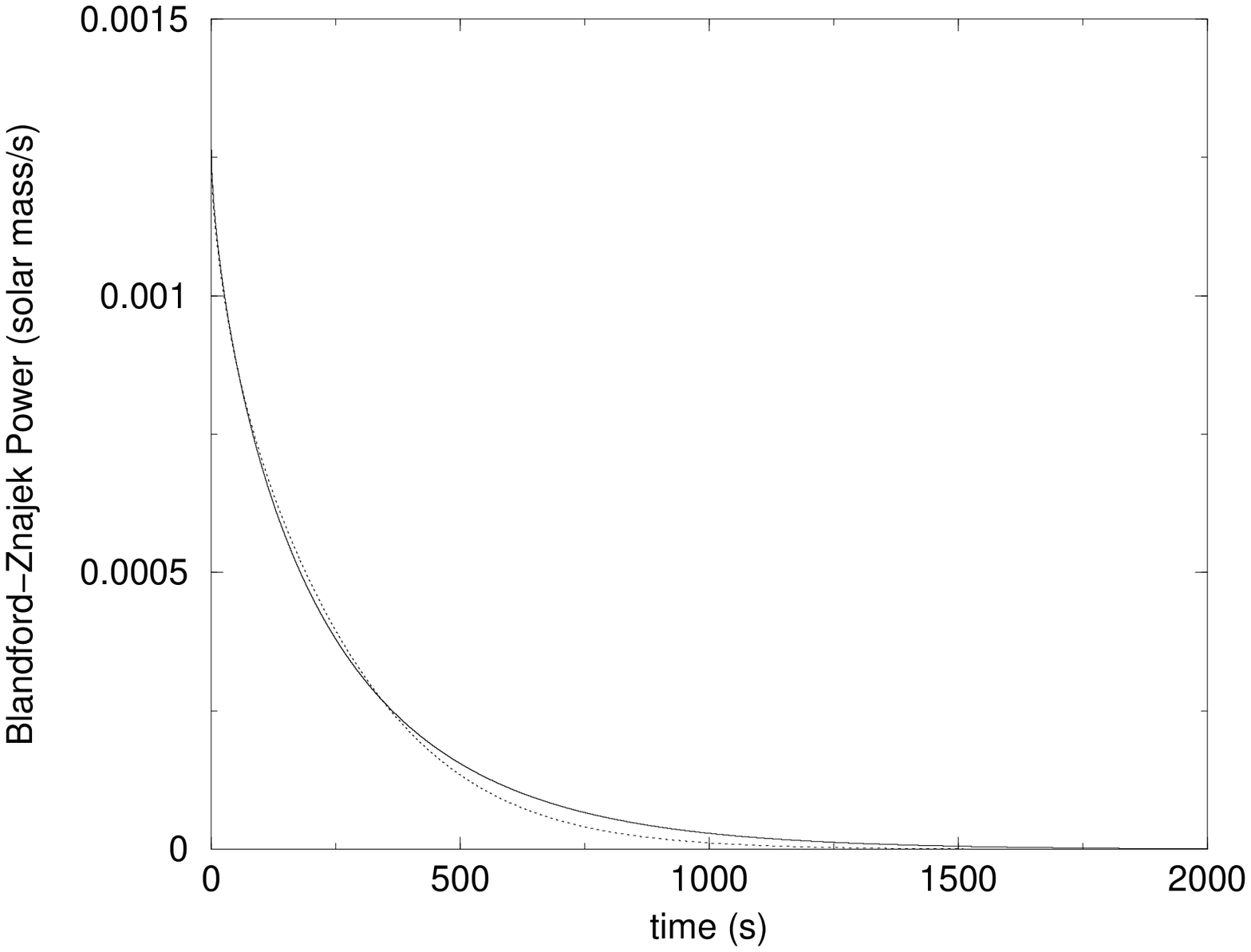, height=11cm}
    \caption{}
    \label{fig3}
  \end{center}
\end{figure}

\end{document}